\documentstyle[aps,multicol,tighten,eqsecnum,graphicx]{revtex}
\begin{document}
\draft
\title{Complete parameterization, and invariance, 
of diffusive quantum trajectories \\
for Markovian open systems}
\author{H.M. Wiseman$^{1,*}$
and L. Di\'osi$^{2,\dagger}$}
\address{$^{1}$School of Science, Griffith University, Brisbane 4111
Australia.}
\address{$^{2}$Research Institute for Particle and Nuclear Physics,
H-1525 Budapest 114, POB 49, Hungary}
\maketitle

\newcommand{\beq}{\begin{equation}}
\newcommand{\eeq}{\end{equation}}
\newcommand{\bqa}{\begin{eqnarray}}
\newcommand{\eqa}{\end{eqnarray}}
\newcommand{\nn}{\nonumber}
\newcommand{\nl}{\nn \\ &&}
\newcommand{\dg}{^\dagger}
\newcommand{\erf}[1]{Eq.~(\ref{#1})}
\newcommand{\smallfrac}[2]{\mbox{$\frac{#1}{#2}$}}
\newcommand{\bra}[1]{\left\langle{#1}\right|}
\newcommand{\ket}[1]{\left|{#1}\right\rangle}
\newcommand{\ip}[2]{\left\langle{#1}\right|\left.{#2}\right\rangle}
\newcommand{\sch}{Schr\"odinger }
\newcommand{\schs}{Schr\"odinger's }
\newcommand{\hei}{Heisenberg }
\newcommand{\heis}{Heisenberg's }
\newcommand{\half}{\smallfrac{1}{2}}
\newcommand{\bl}{{\bigl(}}
\newcommand{\br}{{\bigr)}}
\newcommand{\ito}{It\^o }
\newcommand{\sq}[1]{\left[ {#1} \right]}
\newcommand{\cu}[1]{\left\{ {#1} \right\}}
\newcommand{\ro}[1]{\left( {#1} \right)}
\newcommand{\an}[1]{\left\langle{#1}\right\rangle}
\newcommand{\implies}{\Longrightarrow}
\newcommand{\tr}[1]{{\rm Tr}\sq{ {#1} }}
\newcommand{\st}[1]{\left| {#1} \right|}
\newcommand{\singlecol}{\end{multicols}
                        \vspace{-0.5cm}

\noindent\rule{0.5\textwidth}{0.4pt}\rule{0.4pt}{\baselineskip}
                        \widetext }
\newcommand{\doublecol}{\noindent\hspace{0.5\textwidth}

\rule{0.4pt}{\baselineskip}\rule[\baselineskip]{0.5\textwidth}{0.4pt}
                        \vspace{-1cm}\begin{multicols}{2}\noindent}

\begin{abstract}
The state matrix $\rho$ for an open quantum system with Markovian evolution
obeys a master equation. The
master equation evolution can be {\em unraveled} into stochastic nonlinear
trajectories for a pure state $P$, such that on average $P$
reproduces $\rho$. Here we give for the first time a complete
parameterization of all diffusive unravelings (in which $P$ evolves
continuously but non-differentiably in time). We give an explicit
measurement theory interpretation for these quantum trajectories, in terms of
monitoring the system's environment. We also introduce new
classes of diffusive unravelings that are invariant
under the linear operator transformations under which the master equation is
invariant. We illustrate these invariant unravelings 
by numerical simulations. Finally, we
discuss generalized gauge transformations as a method of connecting
apparently disparate descriptions of the same trajectories by
stochastic \sch equations, and their invariance properties.
\end{abstract}

\begin{multicols}{2}
\narrowtext
\section{Introduction}

It is well known that quantum mixtures differ qualitatively from classical
mixtures. Here a mixed state means one about which we have incomplete
knowledge, as opposed to a pure state, which is one about which we
have maximal knowledge. In the classical case, a mixed state is  described
 by  a probability distribution $\wp$ over phase space. There is a one-to-one
correspondence between this probability distribution $\wp$ and a weighted
ensemble of points in phase space (pure states). In the quantum case,
a mixed state is described by a  density operator, or state matrix
$\rho$. But there is a one-to-many (infinitely many, in fact) mapping
\cite{HugJozWoo93}
from this $\rho$ to a weighted ensemble of state vectors (pure states).
That is to say, there are infinitely many ways to write a given impure
state matrix $\rho$ as a positively weighted sum of projectors.

This difference between quantum and classical systems is also
reflected in the dynamics of open systems. Interaction with an
environment generically causes systems to become mixed. This can
be described by deterministic evolution of the mixed state (the
classical $\wp$ or the quantum $\rho$). For example, this evolution
may be described by a Fokker-Planck equation for $\wp$, or a master
equation for $\rho$. Alternatively, the dynamics can be described by
stochastic trajectories of pure states (the classical point in phase
space or the quantum projector). In the classical case
the two descriptions follow uniquely from each other and they are completely
equivalent. But in the quantum case, the
trajectory equation does not follow uniquely from the
ensemble equation. In fact there are infinitely many different
quantum trajectory equations for a given master equation
\cite{Dio88ab,Gis90}.

It is thus apparent that quantum trajectories have a richer physical
content than classical trajectories. The different quantum trajectory
equations have been called different {\em unravelings} of the master
equation. In some (but not all \cite{WisToo99})
cases, they can be interpreted as arising
from inequivalent  schemes of efficiently monitoring the environment to
which the system is coupled \cite{Wis96a}.
(For classical systems, all such efficient schemes would be equivalent with
one another.)
From this point of view, the statistical
description arises from the probabilistic nature of quantum
measurements. However, it is possible to study the nature of quantum
unravelings without specifying a concrete measurement scheme.

In the present work, we limit our discussion to a particular class of
unravelings, in which the noise in the quantum trajectory is diffusive
in nature. After some preliminaries, we give,
for the first time, a complete parameterization
of such unravelings in Sec.~IV. An explicit formulation in terms of
general quantum measurement is given in Sec.~V. In Sec.~VI we discuss
the notion of invariance of quantum trajectories, using concepts from
Sec.~II. We introduce new classes of invariant unravelings and, in
Sec.~VII, illustrate them for the process of resonance fluorescence.
In Sec.~VIII we discuss the stochastic \sch equation formulation of
quantum trajectories,  and show how seemingly inequivalent equations
may be related by generalized gauge transformations. We conclude with a
summary
and discussion of the relation of our present work to past and future
work in the field.

\section{The Master Equation}

If a quantum system is weakly coupled to an environmental reservoir,
and many modes of the reservoir	are roughly equally affected by
the system, then one can make the Born and Markov approximations
in describing the effect of the	environment on the system
\cite{Gar91}. Tracing over (that is, ignoring) the state of
the environment leads to a Markovian evolution equation	for
the state matrix $\rho$ of the system, known as	a {\em quantum
master equation}. The most general form	of the quantum master
equation which is mathematically valid is the Lindblad form
\cite{Lin76}
\beq
\dot{\rho}={\cal L}\rho\equiv
	-i[H,\rho] + c_k\rho c_k\dg - \half\cu{c_k\dg c_k,\rho}~.
	 \label{genme}
\eeq
Here $\cu{c_k}_{k=1}^{K}$ is the ordered set of Lindblad operators, and
 as in the remainder of this paper,
we are using the Einstein summation convention for repeated indices.
The Lindblad operators couple the system to the
reservoir modes.

The above representation of the evolution superoperator
${\cal L}$ is not unique. We can reduce the ambiguity by requiring that the
operators $1,c_1,c_2,\dots,c_K$ be linearly independent. Then we are left
with the freedom of re-defining the Lindblad operators by an arbitrary
$K\times K$ unitary matrix $T_{kl}$ \cite{GisPer92}:
\beq\label{rotlo}
c_k\rightarrow T_{kl}c_l~. 
\eeq
Here, as in the remainder of this paper,
we are using the Einstein summation convention for repeated indices.
In addition, ${\cal L}$ is invariant under
 c-number shifts of the Lindblad operators, accompanied by a new term in the
Hamiltonian
\beq\label{shiftlo}
c_k\rightarrow c_k+\chi_k~,~~~~
  H\rightarrow H + \smallfrac{i}{2}\ro{\chi_k^\star c_k - {\rm H.c.}}~.
\eeq

The master equation turns pure states $\rho=\ket{\psi}\bra{\psi}\equiv P$
into mixed ones. A related mathematical object is the {\it transition
(mixing) rate operator} \cite{Dio86}
\bqa\label{trrateo}
W &=&{\cal L}P - \cu{P,{\cal L}P}+P{\rm Tr}\sq{P{\cal L}P}
\\ &=& 	 \ro{c_k-\an{c_k}}P\ro{c_k-\an{c_k}}\dg~, 
\eqa
where $\an{\dots}={\rm Tr}[\dots P]$ stands for quantum expectation values.
$W$ is an invariant operator; it does not change with the unitary rotation
(\ref{rotlo}) of the Lindblad operators or the c-number shift
(\ref{shiftlo}). The transition rate operator is
positive semi-definite and orthogonal to the current state $P$, i.e.
$WP=PW=0$. Its trace is the transition (mixing) rate:
\beq\label{trrate}
w\equiv{\rm Tr}W= \ro{ \langle c_k\dg c_k\rangle -
\langle{c_k\dg}\rangle\an{c_k} }~. 
\eeq

\section{Quantum trajectories} \label{Sec3}

In the situation where
a Markovian master equation can	be derived, it is possible (in
principle) to continually measure the state of the environment on
a time scale large compared to the reservoir correlation time but
small compared to the response time of the system. This
effectively continuous measurement is what we will call
``monitoring''.	In such	systems, monitoring the	environment does
not disrupt the system--reservoir coupling and the system will
continue to evolve according to	the master equation if one
ignores	the results of the monitoring.

By contrast, if	one does take note of the results of monitoring
the environment, then the system will no
longer obey the	master equation. Because the system--reservoir
coupling causes	the reservoir to become	entangled with the
system,	measuring the former's state produces information about
the latter's state. That is, the system state is conditioned upon the
result of the measurements. This will tend to undo the increase	in the
mixedness of the system's state	caused by the coupling.

Perfect monitoring of the reservoir requires continual rank-one
projective (i.e. von Neumann) measurements of its state, on the time
scale discussed above. If the system initially has a mixed state, then
its state
will usually be	collapsed towards a pure state.	However	this is
not a process which itself can be described by projective
measurements on	the system, because the	system is not being
directly measured. Rather, the monitoring of the environment
leads to a gradual (on average)	decrease in the	system's entropy.

If the system is initially in a	pure state then, under perfect
monitoring of its environment, it will remain in a pure	state. Then the
effect of the monitoring is to cause the system	to change its pure
state in a stochastic and (in general) nonlinear way. Such evolution
has been called	a {\it quantum trajectory} \cite{Car93b}. It can be
described by a nonlinear stochastic \sch equation (SSE)
\cite
{Gis84,Dio86,Dio88ab,GhiPeaRim90,GisPer92,DalCasMol92,GarParZol92,WisMil93c}:
\beq\label{sse}
\vert\dot\psi\rangle=-iH_\psi\ket{\psi}~+~{\rm noise}_{\psi}~.
\eeq
Here $H_{\psi}$ is a non-Hermitian effective Hamiltonian, and (like the
noise) it depends on $\psi$. The nonlinearity
and stochasticity are present because they are a fundamental part of
measurement in quantum mechanics.

The stochastic average of pure state quantum trajectories still
reproduces the state of the ensemble $\rho$ for each time $t$:
\beq\label{unr}
{\rm E}\sq{~\ket{\psi(t)}\bra{\psi(t)}~}\equiv
{\rm E}[P(t)]=\rho(t)~.
\eeq
Here E denotes an expectation value, or ensemble average with respect
to the noise process in the stochastic \sch equation.
The nonlinear Hamiltonian and the stochastic term in
\erf{sse} must be derived from the above constraint.
Then the stochastic \sch equation is said to {\it unravel}
the master equation \cite{Car93b}. It is now well-known
\cite{QSOSQO96} that there
are many (in fact continuously many) different unravelings for a
given master equation, corresponding to	different
ways of	monitoring the environment.

Any classical noisy trajectory can be approximated to an arbitrary
accuracy by a trajectory consisting of deterministic evolution
punctuated by jumps \cite{Van92}. In the same way, the noise in
the stochastic \sch equation (\ref{sse}) can always be	written	as a
quantum	jump term.  These jumps may range in size from being
infinitesimal, to being	so large that the system state after the
jump is	always orthogonal to that before the jump
\cite{Dio86,BreMilWis95,RigGis96}. 
In this paper we are concerned with the former case.
In the limit of infinitesimal jumps occurring infinitely frequently,
a diffusive unraveling results. As in Brownian motion, the state of the
system evolves continuously but not differentiably in time. For this
reason, these sorts of unravelings have been called continuous
unravelings \cite{WisVac98}, but here we call them diffusive.

Although a stochastic \sch equation is conceptually
the simplest way to define a quantum trajectory, in this work we will
instead	use the stochastic master equation (SME) 
\cite{Dio88ab,Gis89,WisMil93c}.
This has a number of advantages.
First, it is more general in that it
can	describe the purification of an	initially mixed	state. Second, it
is easier to see the relation between the quantum trajectories and the
master equation	which the system still obeys on	average.	Third,
it is invariant under gauge transformations
\beq\label{gauge}
\ket{\psi(t)} \to \exp[i\chi(t)]\ket{\psi(t)},
\eeq
where $\chi(t)$ is an arbitrary real function of time. Such gauge
transformations can radically change the appearance of a stochastic
\sch equation, since $\chi(t)$ may be
stochastic and may be a function of $\ket{\psi}$ itself.
Since unnormalized wave functions may also be used, the
gauge transformation (\ref{gauge}) can be extended for complex functions
$\chi(t)$. We discuss
these points further in Sec.~VIII.

\section{Diffusive Unravelings}

Assuming that the initial state	of the system is pure, the quantum
trajectory for its projector will be described by the SME of the
form $\dot{P}={\cal L}P~+~{\rm noise}$. The drift term, by virtue of
Eq.~(\ref{unr}), assures the consistency with the ensemble evolution
(\ref{genme}). The noise term, which has an expectation value of zero,
we are assuming to be diffusive in nature. It is convenient to
represent such singular Gaussian noise using the \ito{} calculus \cite{Gar85}.
Writing its \ito-differential as
$dX$, we have the following general form of SME:
\beq \label{SMEdX}
d{P}=  dt{\cal L}P + dX~.
\eeq
Here $dX$ is traceless and Hermitian, and depends nonlinearly on the
current pure state $P$. Its expectation value
${\rm E}[dX]$ is zero. Since the SME is assumed to preserve the
purity
of the state, the second moments of $dX$ are constrained by the
identity $dP\equiv d(P^2)=\cu{dP,P}+dPdP$, which must hold for
arbitrary rank-one projectors $P$. Substituting the expression (\ref{SMEdX})
and using the \ito{} rules yields two separate equations:
\beq\label{dX}
dX=\cu{P,dX}~,
\eeq
\beq\label{dXdX}
dXdX={\cal L}Pdt-\cu{P,{\cal L}P}dt~.
\eeq

We can write the general solution of Eq.~(\ref{dX}) in the
simple form:
\beq\label{dXdphi}
dX = \ket{d\varphi}\bra{\psi}+\ket{\psi}\bra{d\varphi},
~~~~\bra{d\varphi}\psi\rangle=0~.
\eeq
Here $\ket{d\varphi}$ is an \ito-differential of zero mean,
orthogonal to the current state $\ket{\psi}$. Its autocorrelation
is constrained: substituting Eq.~(\ref{dXdphi}) into
Eq.~(\ref{dXdX}) yields
\beq\label{dphidphi}
\ket{d\varphi}\bra{d\varphi}=Wdt~,
\eeq
where $W$ is the $P$-dependent transition rate operator (\ref{trrateo}).
Hence we have obtained the general form of diffusive
unravelings in terms of the following SME:
\beq \label{SMEdphi}
d{P}=  dt{\cal L}P + \ket{d\varphi}\bra{\psi}+\ket{\psi}\bra{d\varphi},
\eeq
where the \ito-differential $\ket{d\varphi}$ is orthogonal to $\ket{\psi}$,
has zero mean, and the Hermitian part of its correlation is given by
Eq.~(\ref{dphidphi}).

The careful reader will observe that the non-Hermitian part
$\ket{d\varphi}\ket{d\varphi}$ remains free, expressing the fact that
there are infinitely many pure state diffusive unravelings
of the same Lindblad master equation (\ref{genme}). If the correlation
$\ket{d\varphi}\ket{d\varphi}$  is set to zero then the
noise term is uniquely defined by the Hermitian correlation
in Eq.~(\ref{dphidphi}), and we obtain a unique unraveling
\cite{Dio88c,Per90}. Let us call it the standard one. The standard SME
(\ref{SMEdphi}) follows uniquely from the Lindblad master equation
(\ref{genme}) and it is invariant in a sense that it does not change
with the re-definition $(\ref{rotlo},\ref{shiftlo})$ of the Lindblad
operators. For a long time it has apparently been thought that the 
standard one is the only invariant unraveling 
\cite{Dio88c,GisPer92,RigGis96}. 
In Sec.~VI, however,
we display invariant and non-zero choices for the non-Hermitian
correlation $\ket{d\varphi}\ket{d\varphi}$ of the noise.

Equation~(\ref{SMEdphi}), with the constraints listed below it,
represents diffusive unraveling in complete generality. However, for
many practical purposes, it is useful to have a more explicit
construction. That is we wish to reparametrize the state-vector valued
generalized Wiener noise
$\ket{d\varphi}$ by
complex-number-valued standard Wiener noises. Recall the representation
(\ref{trrateo}) of the transition rate operator $W$ in terms of the
Lindblad operators $c_k$. From Eq.~(\ref{dphidphi}) it is then obvious that
$\ket{d\varphi}$ is spanned by the vectors $\ro{c_k-\an{c_k}}\ket{\psi}$.
We introduce the vector of complex Wiener processes
$\vec{\xi}(t) = \cu{\xi_k(t)}_{k=1}^{K}$ as coefficients:
\beq\label{dphidxi}
\ket{d\varphi}=\ro{c_k-\an{c_k}}\ket{\psi}d\xi_k^*~.
\eeq
Recall that we are using the Einstein summation convention, and note
our notation of using $\xi(t)$ for the Wiener process,
not its time-derivative as in Ref.~\cite{Gar85}.

The mean increments ${\rm E}\sq{d\xi_k}$ vanish, so that ${\rm
E}\sq{\ket{d\varphi}}$ does also. Also,
the above form of $\ket{d\varphi}$ satisfies the constraint (\ref{dphidphi})
provided the Hermitian part of the noises' correlation matrix is the unit
matrix,  while the non-Hermitian part remains free:
\bqa
d\xi_j(t)d\xi_k^*(t) &=& dt	\, \delta_{jk}~, \label{ito1}
\\ d\xi_j(t)d\xi_k(t) &=& dt\, u_{jk}~.	\label{ito2}
\eqa
The $u_{jk}=u_{kj}$ are arbitrary complex numbers subject
only to the condition that the $2K\times 2K$ correlation matrix
of the real vector $\ro{{\rm Re}d\vec{\xi},{\rm Im}d\vec{\xi}}$
\bqa\label{corr}
\frac{dt}{2}
\ro{\begin{array}{cc}{\bf 1}+{\rm Re}[{\bf u}]&{\rm Im}[{\bf u}]\\
                     {\rm Im}[{\bf u}]&{\bf 1}-{\rm Re}[{\bf u}]\end{array}}
\eqa
be nonnegative. It can be shown that the smallest eigenvalue of this
matrix is given in terms of the norm of the complex matrix ${\bf u}$
by $(1-\Vert {\bf u} \Vert)/2$. Thus, this condition is satisfied
if an only if
\beq\label{unorm}
\Vert {\bf u} \Vert\leq1~.
\eeq
With this parameterization we can rewrite the SME (\ref{SMEdphi})
explicitly as
\beq\label{SMEdxi}
dP = {\cal L}Pdt +
\sq{ \ro{c_k-\an{c_k}}Pd\xi_k^* + {\rm H.c.} }~.
\eeq

\section{Measurement Interpretation}

We stated in Sec.~\ref{Sec3} that the master
equation is unraveled if the environment of the system is monitored,
and that the pure state $P$ obeying \erf{SMEdxi} can be interpreted as
the state conditioned on the results of this monitoring.
To see this relationship, it is necessary to consider the theory of
non-projective or indirect measurements (see for example
Ref.~\cite{BraKha92}). Such measurements arise when the system of
interest interacts with a second system, and that second system is
subject to a measurement of the traditional (projective) sort. If the
second system is initially in a pure state, and a rank-one projective
measurement is made on it, then the indirect measurement on the
system can be described by a set of measurement operators
$\Omega_{r}$. Here $r$ labels
the {\it result} of the measurement, and the
operators are constrained only by the completeness relation
\beq
\int d\mu_{0}(r) \Omega_{r}\dg\Omega_{r} = 1~.
\eeq
Here $d\mu_{0}(r)$ is a normalized measure over the space of all $r$.
Let the initial state of the system
be $\rho$.
The measurement operators give both the probability
\beq
d\mu(r) = d\mu_{0}(r) \tr{\rho\Omega_{r}\dg\Omega_{r}},
\eeq
for obtaining a
result in an infinitesimal vicinity of $r$, and the state
\beq
\rho'_{r} = \frac{d\mu_{0}(r) \Omega_{r} \rho \Omega_{r}\dg }{d\mu(r)} \\
=\frac{\Omega_{r} \rho \Omega_{r}\dg }{\tr{\rho\Omega_{r}\dg\Omega_{r}}}
\eeq
conditioned on the result $r$. If the measurement is made but the
result ignored, the new system state is
\beq
\rho' = \int d\mu(r) \rho'_{r} \\
= \int d\mu_{0}(r) \Omega_{r} \rho \Omega_{r}\dg~.
\eeq

In this paper we are concerned with systems that obey the master
equation (\ref{genme}), which is continuous in time. That is to say,
we have to consider repeated indirect measurements, each lasting an
infinitesimal (with respect to the relevant system time scales)
interval of time, such that if one ignores the result,
one obtains
\beq
\rho' = \rho+d\rho = \rho + dt{\cal L}\rho~.
\eeq
In order to obtain the conditioned evolution equation described by
the SME (\ref{SMEdxi}) it turns out that the measurement result $r$ in
any infinitesimal time interval $[t,t+dt)$ must be
described by a vector of complex
numbers $\vec{J}(t)=\cu{J_{k}(t)}_{k=1}^{K}$. As  functions of
time, these are continuous but not differentiable, and we will call
them currents. Explicitly, they are related to the complex Wiener increments
in \erf{SMEdxi} by
\beq \label{defJ}
J_{k}dt = dt\an{u_{kj}c_{j}\dg +c_{k}}+d\xi_{k}~.
\eeq
That is, it is the randomness in the measurement record which provides
the stochasticity in the quantum trajectory.

We can prove this relation between the noise in the quantum
trajectory and the noise in the measurement record by using the
theory of indirect measurements described above.
 We define the measurement operator to be
\beq \label{mo}
\Omega_{\vec{J}}=
1-iHdt-\half c_{k}\dg c_{k}dt + J_{k}^{*}c_{k}dt~.
\eeq
These obey the completeness relation
\beq
\int d\mu_{0}(\vec{J}) \Omega\dg_{\vec{J}} \Omega_{\vec{J}} = 1
\eeq
if we choose $d\mu_{0}(\vec{J})$ to be the measure such that
\bqa
\int d\mu_{0}(\vec{J}) (J_{k}dt) &=& 0~,\\
\int d\mu_{0}(\vec{J}) (J_{j}^{*}dt) (J_{k}dt) &=& \delta_{jk}dt~,\\
\int d\mu_{0}(\vec{J}) (J_{j}dt) (J_{k}dt) &=& u_{jk}dt~.
\eqa
These moments are the same as those of the Wiener
increment $d\vec{\xi}$ as defined
above.

With this assignment of measurement operators $\Omega$ and measure
$d\mu_{0}$ we can easily show that the expected value of the result
$\vec{J}$ is
\beq
{\rm E}[J_{k}] = \int d\mu(\vec{J}) J_{k} = \an{u_{kj}c_{j}\dg
+c_{k}}~.
\eeq
This is consistent with the previous definition in \erf{defJ}.
Furthermore, we can show that the second moments of $\vec{J}dt$ are
(to leading order in $dt$), independent of the system state and can be
calculated using $d\mu_{0}$ rather than $d\mu$. In other words, they
are identical to the statistics of $d\vec{\xi}$ as defined above. This
completes the proof that \erf{defJ} gives the correct probability for
the result $\vec{J}$.

The next step is to derive the conditioned state of the system after
the measurement.
According to the theory of indirect measurements this is given by
\beq
P+dP = \frac{\Omega_{\vec{J}}P\Omega\dg_{\vec{J}}}
{{\rm Tr}\sq{\Omega_{\vec{J}}P\Omega\dg_{\vec{J}}}}~.
\eeq
Expanding to order $dt$ gives
\bqa
dP &=& -\half\cu{c_{k}\dg c_{k},P}dt + J^{*}_{k}dtc_{k}\,P\,c_{l}J_{l}dt \nl{+}
(J^{*}_{k}dt J_{k}-1)\an{c_{j}\dg c_{j}}P dt \nl{+}
 J_{k}^{*}(c_{k}-\an{c_{k}})Pdt + P(c_{k}\dg-\an{c_{k}}\dg) J_{k}dt~.
\eqa
Substituting in the above result (\ref{defJ}) for $\vec{J}$ yields the
required equation (\ref{SMEdxi}). From this it is again obvious that on
average the system obeys the master equation. In the measurement
interpretation this can be derived directly from the nonselective
(ignoring the measurement result) evolution
\beq
d\rho = \int d\mu(\vec{J}) dP
= -P+\int d\mu_{0}(\vec{J}) \Omega_{\vec{J}}P\Omega\dg_{\vec{J}}~.
\eeq

Some insight into the above formalism may be
found by considering an experimentally realisable situation in
quantum optics \cite{Bar90}. For simplicity we consider a system with
 one irreversible term;
that is, $K=1$. For specificity,
say the	system is a two-level atom, with spontaneous emission rate
$\gamma$.
Then $c=\sqrt{\gamma}\sigma$, where $\sigma = \ket{g}\bra{e}$ is
the lowering operator for the cavity.
Say the atom is placed at the focus of a parabolic mirror so that the
emitted light emerges as a beam, and let that beam impinge upon a beam
 splitter of
transmittance $\eta$. Let the transmitted beam be subject to homodyne
detection with a local oscillator of phase (relative to the system)
of $\theta_{1}$. This will yield a real homodyne photocurrent of
\cite{Car93b}
\beq
I_{1}dt =
\sqrt{\gamma\eta}\an{e^{-i\theta_{1}}\sigma+e^{i\theta_{1}}\sigma\dg}+
d\zeta_{1}(t)~,
\eeq
which has been normalized to have a shot-noise spectrum of unity from
the real Wiener process $\zeta_{1}(t)$. Let the reflected beam be
subject to homodyne
detection with a local oscillator of phase $\theta_{2}$, yielding
\beq
I_{2}dt = \sqrt{\gamma(1-\eta)}\an{e^{-i\theta_{2}}\sigma+
e^{i\theta_{2}}\sigma\dg}+d\zeta_{2}(t)~,
\eeq
where $\zeta_{2}(t)$ is an independent real Wiener process.

From
these real currents we can define a `complex current'
\bqa
Jdt &=& \sqrt{\eta_{1}}e^{i\theta_{1}}I_{1}dt + \sqrt{1-\eta}
e^{i\theta_{2}}I_{2}dt \\
&=& \sqrt{\gamma}\an{u \sigma\dg + \sigma}dt + d\xi(t)~,
\eqa
where
\beq
u = \eta e^{2i\theta_{1}} + (1-\eta)e^{2i\theta_{2}}~,
\eeq
and
\beq
d\xi=
\sqrt{\eta}e^{i\theta_{1}}d\zeta_{1}+\sqrt{1-\eta}e^{i\theta_{2}}d\zeta_{2}
\eeq
obeys $d\xi^{*}d\xi = dt$, $d\xi d\xi = udt$. Furthermore, the
conditioned system state can be shown to obey the expected stochastic
master equation \cite{WisMil93a}.

\label{sec:meastint}

\section{Invariant Diffusive Unravelings}

From \erf{SMEdxi} it is obvious that all diffusive unravelings with ${\bf u}$
fixed are invariant
under the shift transformation of \erf{shiftlo}. However,
from Eq.~(\ref{ito2}) we see that the unraveling is
in general not invariant against the unitary rearrangement (\ref{rotlo})
of the Lindblad operators. It is invariant if the
non-Hermitian correlations $u_{jk}$ vanish (the standard diffusive
unraveling). As noted above, it has been guessed that this was the only
invariant unraveling.
We show here that there are further invariant unravelings.

Consider the following non-Hermitian
correlations (\ref{ito2}):
\beq\label{ujk1}
u_{jk}=R\times\an{\ro{c_j-\an{c_j}}\ro{c_k-\an{c_k}}}~.
\eeq
Here $R$ is a complex number constrained only by the fact that its
magnitude must be sufficiently small for the positivity condition
(\ref{unorm}) to be satisfied (note that
$\Vert {\bf u}\Vert$ is invariant).
For a system with unbounded Lindblad operators $c_{k}$, the invariant
 number $R$ would have to depend upon $P$ to ensure this.
An obvious choice would be for $R$ to be real, and to take
the maximally positive (or maximally negative)
value that satisfies \erf{unorm}.
For the special case of finite $N$-dimensional Lindblad operators,
state independent alternatives can also  be chosen, for example
\beq\label{ujk2}
u_{jk}=R\times
\tr{\ro{c_j-N^{-1}{\rm Tr}c_j}\ro{c_k-N^{-1}{\rm Tr}c_k}}~.
\eeq

The above correlations are trivially invariant for the shifts
(\ref{shiftlo}). It is crucial to notice that the coefficient
$u_{jk}$ depends on operator product $c_jc_k$ instead of the
Hermitian versions $c_jc_k\dg$ or $c_j\dg c_k$. This little
difference assures that the SME (\ref{SMEdxi}) will be invariant
for rotations (\ref{rotlo}). To inspect this invariance, observe
that the Lindblad operators and the complex noises appear always in
the same combination $\ket{d\varphi}$ (\ref{dphidxi}).
The mathematical characterization of this Ito-differential is fully
given by the Hermitian correlation (\ref{dphidphi}) which is
invariant (since $W$ invariant itself) and by the non-Hermitian
correlation
\beq\label{dphidphinonherm}
\ket{d\varphi}\otimes\ket{d\varphi}=
u_{jk}^*\ro{c_j-\an{c_j}}\ket{\psi}\otimes
                       \ro{c_k-\an{c_k}}\ket{\psi}dt~. 
\eeq
This latter becomes invariant for unitary rotations (\ref{rotlo})
if we use the non-trivial definitions (\ref{ujk1}) or (\ref{ujk2})
for $u_{jk}$.

It is interesting to note that the choice of correlations
(\ref{ujk1}) implies that the noise process $d\vec\xi(t)$ is no longer
white. That is because the noise correlations depend upon the state of
the system at that time, which obviously depends upon past values of
the noise. Nevertheless, the quantum trajectory defined by \erf{ujk1}
is still Markovian, in that $dP$ depends only upon $P$ at the present
time, and the noise process $d\vec{\xi}$ is still uncorrelated with $P$.
There are other choices of ${\bf u}$ which would make the quantum
trajectory strictly non-Markovian. For example, ${\bf u}$ could depend
upon the past values of the current $\vec{J}$, and in fact there
are very practical	reasons	for	wishing	to consider	such
unravelings \cite{Wis95c}. However we will not be concerned with
this possibility here.

\section{Simulation of Unravelings}

In this section we illustrate various unravelings (invariant and
non-invariant) for a simple but interesting quantum optical system: a
driven, damped two-level atom. The master equation in the interaction
picture is \cite{Gar91}
\beq \label{me2}
\dot{\rho} = \gamma \sigma \rho \sigma\dg -
\frac{\gamma}{2} \cu{\sigma\dg\sigma,\rho}
-i\frac{\Omega}{2}[\sigma_{x},\rho]~.
\eeq
Here $\Omega$ is the driving strength (the Rabi frequency) and damping
occurs through spontaneous emission with the single Lindblad operator
$\sqrt{\gamma}\sigma$ as described in
Sec.~\ref{sec:meastint}.
Physically, all of the light emitted by the atom would have to be
collected and measured by homodyne-like measurements (as described in
Sec.~\ref{sec:meastint}) in order for a diffusive SSE to describe the
conditioned dynamics of the system.

Since there is only one Lindblad operator, all diffusive unravelings
are defined by  just one complex parameter $u$ satisfying $|u|\leq 1$.
In all cases illustrated below the damping rate $\gamma$ is set to
unity, and the driving rate $\Omega=10\gamma$. We plot the state
using the three components of the Bloch vector $x,y,z$, defined by
\beq
P = \half\ro{1+x\sigma_{x}+y\sigma_{y}+z\sigma_{z}}.
\eeq
Here $\sigma_{x},\sigma_{y},\sigma_{z}$ are the usual Pauli pseudospin
matrices where the up ($\ket{e}$) and down ($\ket{g}$) states
are the $\sigma_{z}$ eigenstates with eigenvalues $+1$ and $-1$,
respectively.
One of the Bloch vector
components is redundant since for a pure state $P^{2}=P$ which
implies
\beq
x^{2}+y^{2}+z^{2} = 1.
\eeq
Nevertheless, it is easiest to interpret the results if we plot all
three. The driving causes the Bloch vector to rotate around the $x$ axis,
while the damping causes it to decay towards the down state ($z=-1$).
In all cases the initial state is a
positive $x$ eigenstate. We emphasize that the ensemble average behaviour is
the same for all unravelings. In particular, after transients have
decayed the system on average reproduces the stationary solution of the
master equation. In the high driving limit,
the steady state of the master equation is
close to a completely mixed state, so that $x$ (exactly), and
$y$ and $z$ (approximately) average to zero.

We begin with two non-invariant unravelings, $u=1$ and $u=-1$. These
correspond to homodyne detection as described in
Sec.~\ref{sec:meastint}. The noise correlations (\ref{corr}) 
degenerate and a standard {\it real} white noise $\zeta$ 
will drive the quantum trajectories. For $u=1$, 
the current (\ref{defJ}) becomes real, with average
\beq \label{xhomJ}
{\rm E}[J(t)] = \sqrt{\gamma}\,\an{\sigma_{x}}~.
\eeq
and noise 
$\xi=\xi^*=\zeta$. The 
SME (\ref{SMEdxi}) can be written as follows:
\beq\label{SMEu1}
dP = {\cal L}Pdt 
+ \smallfrac{i}{2}[\sigma_{y},P]d\zeta
+ \smallfrac{1}{2}\{\sigma_{x}-\an{\sigma_{x}},P\}d\zeta~.
\eeq
The third term on the SME's RHS corresponds to a 
measurement of the $x$ quadrature of the system, as reflected in the 
expectation value of the current (\ref{xhomJ}). However, there is a 
second white noise term  on the RHS, corresponding to
a noisy Hamiltonian. It can be 
interpreted as an additional (non-Heisenberg) 
back-action caused by the monitoring, as if the current noise  was 
being  fed-back to alter the system dynamics. The presence of two 
(correlated) noise terms in the SME, one describing Heisenberg 
back-action and one describing Hamiltonian noise, is a generic feature 
of unravelings with a non-Hermitian Lindblad operator (in this case, 
$\sigma$).

\begin{figure}
\includegraphics[width=0.45\textwidth]{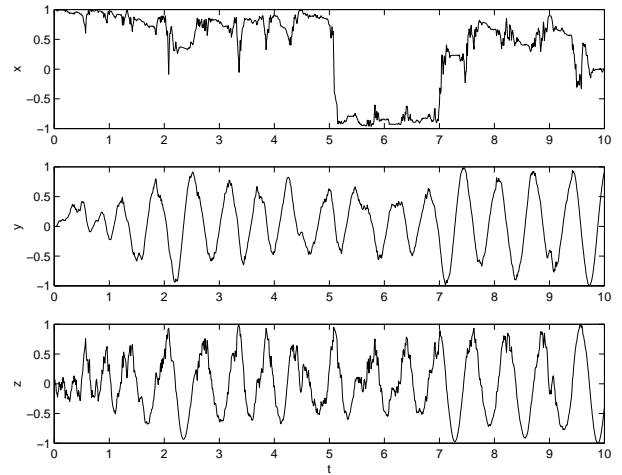}
\vspace{0.2cm}
\caption{\narrowtext Non-invariant unraveling with $u=1$ (homodyne
measurement of $x$ quadrature). }
\protect\label{homoX}
\end{figure}
Fig.~\ref{homoX} shows the conditioned
evolution for $u=1$. Monitoring the $x$ quadrature tends to
make $x$ well-defined (i.e. close to the $\sigma_{x}$ eigenvalues of
$\pm 1$). However, it is certainly not perfect in this respect, as
large oscillations in $y$ and $z$ due to the rotation around the $x$
axis at rate $\Omega$ are still evident.

\begin{figure}
\includegraphics[width=0.45\textwidth]{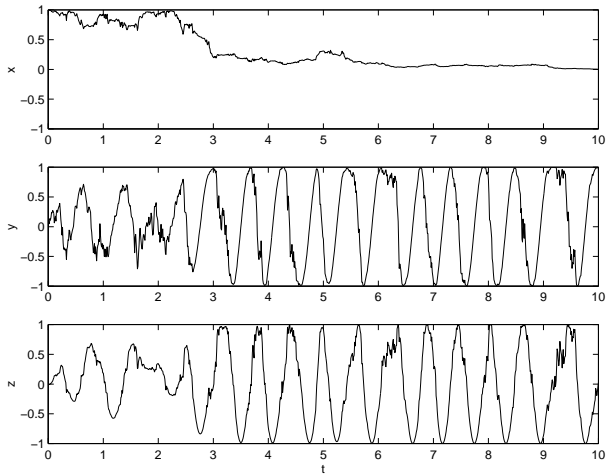}
\vspace{0.2cm}
\caption{\narrowtext Non-invariant unraveling with $u=-1$ (homodyne
measurement of $y$ quadrature).}
\protect\label{homoY}
\end{figure}
Figure \ref{homoY} shows the case $u=-1$, which corresponds to a
homodyne measurement of the $y$
quadrature of the system, so
\beq
{\rm E}[J(t)] = i\sqrt{\gamma}\,\an{\sigma_{y}}~.
\eeq
The measurement tries to make $y$ well-defined,
but fails because the Rabi cycling rotates $y$
into $z$ and so on. Nevertheless, $x$ is forced towards zero, where it
stays.

\begin{figure}
\includegraphics[width=0.45\textwidth]{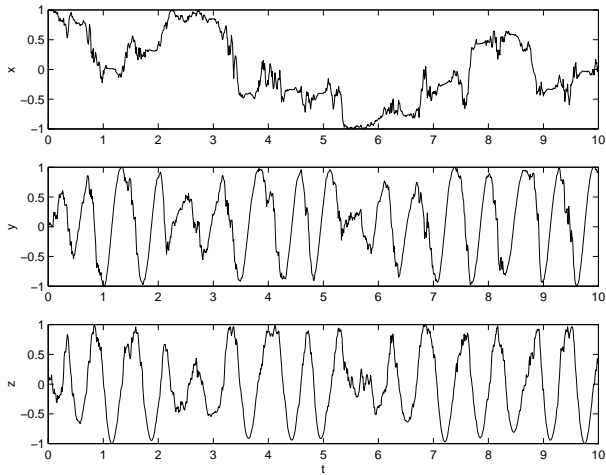}
\vspace{0.2cm}
\caption{\narrowtext Invariant unraveling with $u=0$ (heterodyne
detection).}
\protect\label{het}
\end{figure}
The standard invariant unraveling, or ``quantum state diffusion''
\cite{GisPer92} case
of $u=0$ is shown in Fig.~\ref{het}. This could be realized by heterodyne
detection \cite{WisMil93c}, which is like homodyne detection but with a far-detuned local
oscillator. This ensures that both quadratures are sampled equally.
The current $J$ (which is the complex Fourier transform of the
physical photocurrent) has a mean
\beq
{\rm E}[J(t)] = \sqrt{\gamma}\,\an{\sigma}~.
\eeq
The resultant evolution is intermediate between that of $u=1$ and
$u=-1$: both $\sigma_x$ and $\sigma_y$ are being equally monitored,
and the result is controlling a certain dynamical feed-back 
\cite{DioGisHalPer95}.

Next we plot the simulation of the first of our new invariant unravelings. We
choose $u$ to be given by \erf{ujk1}, where $R$ is chosen to be the
positive number such that $|u|=1$ [the alternative, \erf{ujk2}, simply gives
$u=0$ again]. In this case we find $u=-\an{\sigma}/\an{\sigma\dg}$,
so that
\beq
{\rm E}[J(t)] = 0~.
\eeq
That is, the current consists purely of white noise. This is not 
because the monitoring no longer gives any information about the 
system; it is merely because the 
the measured quadrature of $\sigma$ 
is always chosen to be the one that has an instantaneous mean value 
of zero.

\begin{figure}
\includegraphics[width=0.45\textwidth]{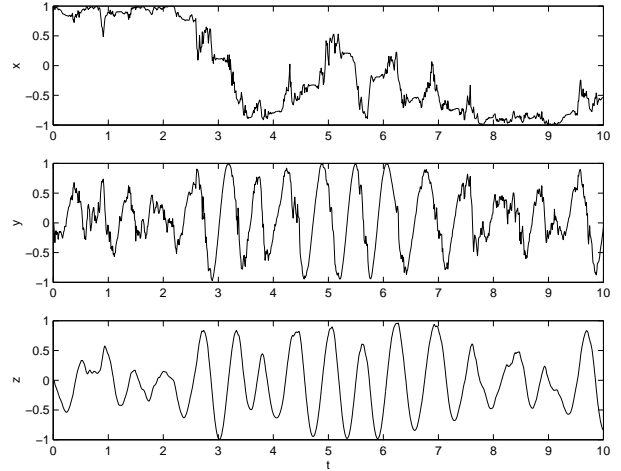}
\vspace{0.2cm}
\caption{\narrowtext Invariant unraveling with $u$ given by \erf{ujk1}
with maximally positive $R$.}
\protect\label{Diosi0}
\end{figure}
 The resultant
trajectories are shown in Fig.~\ref{Diosi0}. It appears that the behaviour is
qualitatively similar to that of $u=0$. However, note
that unlike any of the other cases, the evolution of $z$ is
differentiable. This can be proven analytically, since in this case
$z$ obeys
\beq
\dot{z}=-\gamma z -\gamma +\Omega y~,
\eeq
which is exactly the same as that obtained from the master equation
(\ref{me2}). The evolution of $z$ is of course still stochastic
 because of the coupling to $y$. This feature is a special case of a
 general phenomenon noted in Ref.~\cite{DorNie98}, namely that the
 noise correlations may be chosen
 to make the evolution of any given
 operator average  smooth.

The final plot, Fig.~\ref{Diosi1}, is for the new invariant unraveling with
$u$ given by \erf{ujk1}, where now $R$ is chosen to be the
negative number such that $|u|=1$. This gives
$u=\an{\sigma}/\an{\sigma\dg}$ so that
\beq
{\rm E}[J(t)] = \sqrt{\gamma}\,2\an{\sigma}~.
\eeq
In this case the behaviour appears
qualitatively similar to that of $u=-1$, in that $x$ is forced to zero.
Moreover, the two unravelings becomes equivalent
once $x$ reaches $0$, since then $\an{\sigma_{y}}=2\an{\sigma}$. This
is interesting, in that an invariant unraveling and a non-invariant
unraveling are actually identical in the steady state.
\begin{figure}
\includegraphics[width=0.45\textwidth]{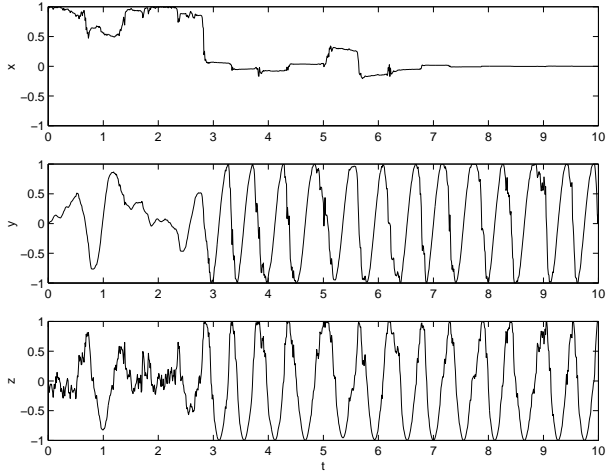}
\vspace{0.2cm}
\caption{\narrowtext Invariant unraveling with $u$ given by \erf{ujk1}
with maximally negative $R$.}
\protect\label{Diosi1}
\end{figure}

\section{Stochastic Schr\"odinger Equations}

We can rewrite the general SME (\ref{SMEdphi}) into the form of a
SSE (\ref{sse}):
\beq\label{SSEdxi}
d\ket{\psi}=-iH_\psi dt\ket{\psi}
	+ \ro{c_k-\an{c_k}}d\xi_k^*\ket{\psi}~. 
\eeq
Using the identity
$dP=(d\ket{\psi})\bra{\psi}+\ket{\psi}(d\bra{\psi})+d\ket{\psi}d\bra{\psi}$,
one can
inspect that the above SSE leads to the SME if the nonlinear
Hamiltonian is chosen as it follows:
\bqa\label{Hnonlin}
-iH_\psi &=&{\cal L}\ket{\psi}\bra{\psi}+\half w \label{Hinvar}\\
	&=&-iH-\half\ro{c\dg_k c_k - 2\an{c_k}^* c_k
	      +\an{c_{k}}^{*}\an{c_k}}~. 
\eqa
It is remarkable that $H_\psi$, apart from irrelevant c-number terms, does
not depend on the particular representation of the master equation.
This is obvious from \erf{Hinvar}.
Let us re-group its terms in the following ways:
\bqa\label{Hnonlin1}
H_\psi=&&
H +\smallfrac{1}{2}\ro{i\an{c_k}^* c_k+{\rm H.c.}}\nonumber\\ 
&&-\smallfrac{i}{2}\ro{c_k-\an{c_k}}\dg\ro{c_k-\an{c_k}}~. 
\eqa
The second term on the r.h.s. is a nonlinear Hermitian term, a kind of
mean-field potential. It is invariant under the rotation
(\ref{rotlo}) and its change under the shift (\ref{shiftlo}) exactly
cancels the change in $H$. The third term is nonlinear and non-Hermitian,
it is responsible for the localization of the wavefunction as a result
of continuous monitoring. It is obviously invariant under both
transformations
(\ref{rotlo}) and (\ref{shiftlo}).

The invariance of $H_{\psi}$ is a nice feature of the  SSE
(\ref{SSEdxi}), but it must be emphasized that it
is not a required property of a SSE.
That is because the mapping from SME to SSE is not unique, since a
gauge (global phase) transformation
\beq
\ket{\psi(t)} \to e^{i\chi(t)}\ket{\psi(t)} = \ket{\phi(t)}
\eeq
does not change the projector
$P=\ket{\psi}\bra{\psi}=\ket{\phi}\bra{\phi}$.
For a non-stochastic \sch equation, this transformation
will cause only trivial changes to its equation of motion;
it will simply add a c-number to the
Hamiltonian. However for a SSE it can radically change the equation.
This has been noted before (see for example Ref.~\cite{DorNie00a}),
but for completeness we show it explicitly.

Consider the case $K=1$ so that we have
one complex noise process $d\xi$, and one Lindblad operator $c$ which
we will take to be Hermitian and write as $x$. Then the SSE
(\ref{SSEdxi}) becomes
\beq  \label{xSSE}
d\ket{\psi}=\sq{-iH-\half\ro{x-\an{x}}^{2}} dt\ket{\psi}
	+ \ro{x-\an{x}}d\xi^*\ket{\psi}~. 
\eeq
Now let the global phase obey the equation
\beq
d\chi = f(t)d\xi^{*} + {\rm c.c.}~,
\eeq
where $f(t)$ is an arbitrary smooth function of time. Then
\beq
\ket{{\phi}}+d\ket{{\phi}} =
\ro{ 1+id\chi-\half d\chi d\chi}e^{i\chi(t)}
\ro{\ket{\psi}+d\ket{\psi}}~.
\eeq
The resultant equation
for $\ket{\phi}$ is
\bqa
d\ket{\phi} &=& \sq{-iH-{\rm Re}\ro{f^{2}u^{*}+|f|^{2}} }dt\ket{\phi}\nl{-}
\half\ro{x-\an{x}}\ro{x-\an{x}+ifu^{*}+if^{*}}
 dt\ket{\phi} \nl{+} \sq{\ro{x-\an{x}+if}d\xi^*+if^{*}d\xi}\ket{\phi}~.
\eqa
Clearly the deterministic part of this is
different from \erf{xSSE}, and not invariant.

In this case the transformed equation seems far less appealing in form
than the original. However, one can find different forms of the general
SSE (\ref{SSEdxi}) which, while not having an invariant $H_{\psi}$,
have other attractions.
In particular, consider the non-unitary gauge transformation defined by
\beq
\ket{\psi(t)} \to e^{i\chi(t)}\ket{\psi(t)} = \ket{\bar\phi(t)}~,
\eeq
where $\chi$ is the complex function obeying
\beq
id\chi = \an{c_{j}}d\xi^{*}_{j} - \half
u_{jk}^{*}\an{c_{j}}\an{c_{k}}dt +\half\an{c_{j}}^{*}\an{c_{j}}dt~.
\eeq
The state $\ket{\bar{\phi}}$ is not normalized,
which is why it is indicated with an overbar. This is not important,
however, as the projector can be defined as
$P=\ket{\bar\phi}\bra{\bar\phi}/\ip{\bar\phi}{\bar\phi}$, and this
will still obey the SME.

Following the same method as above, one finds that $\ket{\bar{\phi}}$
obeys the SSE
\bqa \label{SSE2}
d\ket{\bar\phi} &=& dt\ro{-iH -\half c\dg_{k}c_{k} +
J^{*}_{k}c_{k}}\ket{\bar\phi} 
\eqa
where it is the normalized state $P$
that is used to define the quantum averages in the expression
(\ref{defJ}) for the currents $J_{k}$.
This SSE has a number of nice features. First, it is as
simple as the invariant version (\ref{SSEdxi}). Second,
 it very clearly shows the conditioning of the state on the measurement
result,
and is closely related to the measurement operators (\ref{mo}).
Third, it is an easy form to use for numerical calculation
(and was in fact used  for the simulations in Sec.~VII).

\section{Discussion}

In this paper we have presented new results, and corrected and
clarified old results, pertaining to diffusive unravelings of
Markovian quantum system dynamics.

First, we have given for the first time the most general form of
diffusive quantum unravelings, in Sec.~IV. 
While Gisin restricted his early work \cite{Gis90} for
the two-dimensional special case,
there have been recent
publications which claim to do much the same thing, but in fact do not do
so. For example, the recent work of Adler \cite{Adl00} is restricted
to Lindblad operators that are Hermitian. Dorsselaer and Nienhuis
\cite{DorNie00b} give a general SSE for a master equation with one
Wiener noise term $d\xi$, but for generalizing to the
set $\cu{d\xi_{k}}_{k=1}^{K}$
they say ``we have to assume that the different $d\xi_{k}$ are
uncorrelated''. As we have shown here, this is not a necessary
assumption; $d\xi_{j}d\xi_{k}=u_{jk}dt$ may be nonzero for $j\neq
k$. The non-Hermitian correlation matrix ${\bf u}$ was introduced by one of us
and Vaccaro \cite{WisVac98}, where it was stated that
 $\forall j,k \, |u_{jk}|\leq 1$.
While this is a necessary condition, it is not sufficient.
Here we have shown that a necessary and sufficient condition is
that the norm $\Vert {\bf u} \Vert$  be bounded above by unity.

Second, we have given for the first time the relation between the
most general unraveling, as parameterized by ${\bf u}$, and quantum
measurement theory. The measurement results which condition the
system and so unravel the master equation are $K$ complex
``currents'' (continuous functions of time) given explicitly in Sec.~V.
The measurement interpretation of the diffusive unravelings is
significant in that it means that the ensembles of pure states
resulting from the unraveling can be {\em physically realized}.
Physical realizability was proposed in Ref.~\cite{WisVac98}
as one of the requirements (along with
maximal robustness) for finding the ``most natural'' ensemble of pure
states to represent the mixed state of an open quantum system. Our
present work is significant for this program (continued in
Ref.~\cite{WisBra00}) of investigating decoherence in that gives a
simple boundary ($\Vert {\bf u}\Vert \leq 1$) to the parameter space
$\cu{\bf u}$ of all
diffusive unravelings.

Third, we have corrected the long-standing conception 
\cite{Dio88c,GisPer92,RigGis96}
that the only
invariant unraveling is the standard diffusive unraveling
with ${\bf u}={\bf 0}$. Here invariant means invariance under the
linear transformations of the Lindblad and Hamiltonian operators which
leave the master equation invariant. 
In Sec.~VI we  constructed some
explicit examples of
invariant unravelings with non-zero ${\bf u}$. The most natural ones
have a state-dependent ${\bf u}$ satisfying $\Vert {\bf u} \Vert =
1$. We illustrated two of these, along with the standard invariant unraveling
and some non-invariant unravelings, by numerical simulations of
resonance fluorescence in
Sec.VII. One of the new schemes produced quite distinctive
dynamics for the atom,
which ties into recent work on minimizing statistical errors in
ensemble average simulations using quantum
trajectories \cite{DorNie98}.
``Quantum state diffusion theory'' \cite{GisPer92} suggests the standard
unraveling as the ``most natural''.
The existence of non-standard invariant unravelings calls for
additional arguments. Recently one of us and Kiefer 
\cite{DioKie00} have applied
robustness criteria (different from those in Ref.~\cite{WisVac98})  
to open system unravelings and have approved the standard one.

Fourth, in Sec.~VIII
we have clarified the notion of invariance in the context of
stochastic \sch equations (SSEs) as a way of representing quantum
trajectories. It turns out that gauge transformations can radically
alter the structure of a given SSE. In particular, the invariance
of the standard unraveling is completely destroyed in a generic
gauge.  
This suggests that the best way
conceptually to represent quantum trajectories
is as a stochastic master equation
(SME) for the state projector rather than a SSE for the state vector.
Gauge freedom may, on the other
hand, allow for equivalent SSEs with different attractions, as 
demonstrated.  
The SSEs may be given priority over the SME in numerical calculations, 
but it must be ensured that 
all theoretical claims 
do not rely on non-gauge-invariant properties. 

\acknowledgments
HMW would like to thank Tony O'Connor for mathematical assistance.

\end{multicols}


\begin{references}

\bibitem[*]{email1}Electronic address:
h.wiseman@gu.edu.au

\bibitem[\dagger]{email2}Electronic address:
diosi@rmki.kfki.hu

\bibitem{HugJozWoo93}
L.P. Hughstone, R. Jozsa and W.K. Wootters,
Phys. Lett. {\bf 183A}, 14 (1993).

\bibitem{Dio88ab}
L. Di\'osi,
Phys. Lett. {\bf 129A}, 419 (1988); {\bf 132A}, 233 (1988).

\bibitem{Gis90}
N. Gisin,
Helv. Phys. Acta {\bf 63}, 929 (1990).

\bibitem{WisToo99}
H.M. Wiseman and G.E. Toombes,
Phys. Rev. A {\bf 60}, 2474 (1999).

\bibitem{Wis96a}
H.M. Wiseman,
Quantum Semiclass. Opt {\bf 8}, 205 (1996).

\bibitem{Gar91}
C.W. Gardiner,
{\em Quantum Noise}
(Springer, Berlin, 1991).

\bibitem{Lin76}
G. Lindblad,
Commun. Math. Phys. {\bf 48}, 199 (1976).

\bibitem{GisPer92}
N. Gisin and I.C. Percival,
J. Phys. A, {\bf 25}, 5677 (1992).

\bibitem{Dio86}
L. Di\'osi,
Phys. Lett. {\bf 114A}, 451 (1986).

\bibitem{Car93b}
H.J. Carmichael,
{\em An Open Systems Approach to Quantum Optics}
(Springer-Verlag, Berlin, 1993).

\bibitem{Gis84}
N. Gisin,
Phys. Rev. Lett. {\bf 52}, 1657 (1984).

\bibitem{GhiPeaRim90}
G.C. Ghirardi, Ph. Pearle and A. Rimini,
Phys. Rev. {\bf A42}, 78 (1990).

\bibitem{DalCasMol92}
J. Dalibard, Y. Castin and K. M\o lmer,
Phys. Rev. Lett. {\bf 68}, 580 (1992).

\bibitem{GarParZol92}
C.W. Gardiner, A.S. Parkins, and P. Zoller,
Phys. Rev. A {\bf 46}, 4363 (1992).

\bibitem{WisMil93c} 
H.M. Wiseman and G.J. Milburn,
Phys. Rev. A {\bf 47}, 1652  (1993).

\bibitem{QSOSQO96}
Quant. Semiclass. Opt. {\bf 8} (1) (1996),
special issue on ``Stochastic quantum optics",
edited by H.J. Carmichael.

\bibitem{Van92}
N.G. van Kampen,
{\em Stochastic Processes in Physics and Chemistry 2e}
(North Holland, 1992, Amsterdam).

\bibitem{BreMilWis95} 
J.K. Breslin, G.J. Milburn and H.M. Wiseman,
Phys. Rev. Lett. {\bf 74}, 4827 (1995).

\bibitem{RigGis96}
M.~Rigo and N.~Gisin,
p.~255 of Ref.~\cite{QSOSQO96} (1996).

\bibitem{Gis89}
N. Gisin,
Helv. Phys. Acta {\bf 62}, 363 (1989).

\bibitem{WisVac98}
	H.M. Wiseman and J.A. Vacarro,
	Phys. Lett. A {\bf 250}, 241 (1998).

\bibitem{Gar85}
C.W. Gardiner,
{\em Handbook of Stochastic Methods}
(Spring\-er, Berlin, 1985).

\bibitem{Dio88c}
L. Di\'osi, J. Phys. A {\bf 21}, 2885 (1988).

\bibitem{Per90}
I.C. Percival, London University Reports No. QMW DYN 90-5
and QMW DYN 90-6, 1990 (unpublished).

\bibitem{BraKha92}
V.B. Braginsky and F.Y. Khalili,
{\em Quantum Measurement}
(Cambridge University Press, Cambridge, 1992).

\bibitem{Bar90}
A. Barchielli,
Quantum Opt. {\bf 2}, 423 (1990).

\bibitem{WisMil93a} 
H.M. Wiseman and G.J. Milburn,
Phys. Rev. A {\bf 47}, 642 (1993).

\bibitem{Wis95c}
H.M. Wiseman,
Phys. Rev. Lett. {\bf 75}, 4587 (1995).

\bibitem{DioGisHalPer95}
L. Di\'osi, N. Gisin, J. Halliwell and I.C. Percival
Phys. Rev. Lett. {\bf 74}, 203 (1995).

\bibitem{DorNie98}
F.E. van Dorsselaer and G. Nienhuis,
Eur. Phys. J. D {\bf 2}, 175 (1998).

\bibitem{DorNie00a}
F.E. van Dorsselaer and G. Nienhuis,
J. Opt. B {\bf 2}, R25 (2000).

\bibitem{Adl00}
S.L. Adler,
Phys. Lett. A {\bf 265}, 58 (2000).

\bibitem{DorNie00b}
F.E. van Dorsselaer and G. Nienhuis,
J. Opt. B {\bf 2}, L5 (2000).

\bibitem{WisBra00}
H.M. Wiseman and Z. Brady,
Phys. Rev. A {\bf 62}, 023805 (2000).

\bibitem{DioKie00}
L. Di\'osi and C. Kiefer,
Phys. Rev. Lett. {\bf 85}, 3552 (2000). 


\end{references}
\end{document}